\documentclass{ws-p8-50x6-00}

\usepackage{graphicx}
\usepackage{amsmath}
\usepackage{latexsym}

\def\NCA{{\em Nuovo Cimento} A }

\def\NPB{{\em Nucl. Phys.} B }
\def\NPA{{\em Nucl. Phys.} A }
\def\PLB{{\em Phys. Lett.}  B }

\def\PRD{{\em Phys. Rev.} D }
\def\PRC{{\em Phys. Rev.} C }
\def\ZPC{{\em Z. Phys.} C }
\def\JPA{{\em J. Phys.} A }
\def\EPJC{{\em Eur. Phys. J.} C }
\def\slashchar#1{\setbox0=\hbox{$#1$}
   \dimen0=\wd0 \setbox1=\hbox{/} \dimen1=\wd1
   \ifdim\dimen0>\dimen1 \rlap{\hbox to \dimen0{\hfil/\hfil}} #1
   \else  \rlap{\hbox to \dimen1{\hfil$#1$\hfil}} / \fi}
\newcommand{\bjlim}{{\stackrel{\scriptstyle{\rm Bj}}
{\textstyle\longrightarrow}}}

\def\be{\begin{equation}}
\def\ee{\end{equation}}
\def\bea{\begin{eqnarray}}
\def\eea{\end{eqnarray}}
\begin{document}

%To Prof Nick Karayiannis -- do read this:-
%If needed the word of Chapter~1, you can type in at the 
%\title{}. The words will be in caps and lowercase. 
%For chapter title can be in all caps or in caps and lowercase.
%It is up to the author to type for the case sensitive but 
%all articles must be in the same style. 
%But mostly for Review Volume are without this Chapter~1.
%Thank you
%Jessie   13/4/2000

\title{PARTON DISTRIBUTIONS FOR THE PION IN A CHIRAL QUARK
MODEL\footnote{Talk given at the Workshop on " Lepton Scattering,
Hadrons and QCD `` March 26 -- April 6, 2001.  ; Adelaide ( Australia
)}}

\author{E. RUIZ ARRIOLA}

\address{Departamento de Fisica Moderna, Universidad de Granada,
\\ 18071 GRANADA, Spain\\E-mail: earriola@ugr.es} 


\maketitle\abstracts{Parton distributions for the pion are studied in
a chiral quark model characterized by a quark propagator for which a
spectral representation is assumed.  Electromagnetic and chiral
symmetry constraints are imposed through the relevant Ward-Takahashi
identities for flavoured vertex functions.  Finiteness of the theory,
requires the spectral function to be non-positive
definite. Straightforward calculation yields the result that the pion
structure function becomes one in the chiral limit, regardless of the
details of the spectral function. LO and NLO evolution provide a
satisfactory description of phenomenological parameterizations of the
valence distribution functions but fails to describe gluon and sea
distributions.}

\section{Introduction}
Deep inelastic scattering (DIS) is regarded as one of the traditional
ways to unveil the quark substructure of hadrons \cite{Ro90}. The
hadronic tensor for electro-production on a (unpolarized) hadron is
defined as
\begin{eqnarray}
W_{\mu\nu}(p,q) = \frac{1}{4\pi} \int d^4 \xi 
e^{i q \cdot \xi} \langle p \left| \left[ J^{\rm em}_\mu (\xi) ,
J^{\rm em}_\nu  (0)
\right] \right| p \rangle 
\label{tensor} 
\end{eqnarray} 
where $p$ is the hadron momentum, $J^{\rm em}_\mu (\xi )= \bar q (\xi)
\hat Q \gamma_\mu q (\xi) $ the electromagnetic current with $\hat
Q={\rm diag} (e_u, e_d, e_s) $ the quark charge matrix and $q$ the
photon momentum transfer. Introducing the relativistically invariant
Euclidean momentum transfer $Q^2=-q^2 $ and the Bjorken variable $x=
Q^2 / 2 p \cdot q $, in the Bjorken limit ( $Q^2 \to \infty$ with $x $
fixed), Eq.(\ref{tensor}) can be expressed in terms of the quark and
antiquark distribution functions, $q_i (x)$ and $\bar q_i(x)$
\begin{eqnarray}
W_{\mu\nu} \bjlim F(x) \left[ - g_{\mu\nu} + \frac{q_\mu q_\nu}{ q^2 }
-\frac{1}{q^2} \left( p_\mu - \frac{q_\mu}{ 2 x } \right) \left( p_\nu -
\frac{q_\nu}{2 x } \right) \right]
\end{eqnarray} 
with $ F(x) = \sum_i e_i^2 \left[ q_i (x) + {\bar q}_i (x) \right] /2
$ as a consequence of Lorentz gauge invariance, scaling and the
underlying spin $1/2$ nature of quarks.  
Logarithmic scaling violations due to perturbative QCD radiative
corrections \cite{AP77}, relate in a linear fashion the structure
functions at a given reference scale, $Q_0^2$, to the scale of
interest, $Q^2$,
\begin{eqnarray} 
F (x,Q^2 ) = U(Q^2, Q^2_0 ) F (x,Q_0^2 ).  
\label{evol} 
\end{eqnarray} 
Here, $U(Q^2,Q_0^2)$ is a linear matrix operator, fulfilling
renormalization group properties $ U(Q_1^2 , Q_2^2 ) U(Q_2^2 , Q_3^2 )
= U(Q_1^2 , Q_3^2 ) $ and $ U(Q^2,Q^2)=1 $ and which can be evaluated
for high $Q^2$. Thus, one is left with the calculation of structure
functions at a certain scale $Q_0$ (the so called initial condition,
$F(x,Q_0^2 )$) and the assumption that the scale is high enough to
make perturbative matching meaningful.

In spite of the wealth of experimental and theoretical information
\cite{GRV95,MRST98,GRV98,La00,SMRS92}, first principle calculations of
the initial condition, $F(x,Q_0^2)$, remain, so far, elusive. Thus,
theoretical quark models of hadrons and quark-hadron duality
assumptions are used. This obviously raises embarrassing questions on
confinement, because the intermediate states in the commutator of
Eq.(\ref{tensor}) should be physical hadron states and, instead, one
uses the virtual Compton amplitude \cite{JA85}
\begin{eqnarray}
T_{\mu \nu} (p,q)= i \int d^4 \xi e^{i q \cdot \xi} \langle p \left|
T\left\{ J_\mu^{\rm em} (\xi) J_\nu^{\rm em} (0) \right\} \right| p
\rangle
\label{virtual-compton} 
\end{eqnarray}  
with $W_{\mu\nu} = 1/(2\pi){\rm Im} T_{\mu\nu} $. Introducing
 light-cone (LC) coordinates $x=(x^+,x^-, \vec x_\perp)$ with
 $x^{\pm}=x^0 \pm x^3 $ one gets the following expression for the
 initial condition 
\begin{eqnarray}
F (x, Q_0^2 ) = -\frac{i}{4 \pi} \int \frac{ d k^- d^2 \vec k_\perp}{
(2\pi)^3 } {\rm tr} \left[ \hat Q^2 \gamma^+ \chi(p,k) \right]
\Big|_{k^+ = p^+= mx}
\label{quark-target} 
\end{eqnarray} 
where the forward quark-target scattering amplitude is defined
\begin{eqnarray}
\chi(p,k) = -i \int d^4 \xi \, e^{i \xi \cdot k} \langle p | T \{ q(\xi) \bar
q(0) \} | p \rangle 
\label{q-t} 
\end{eqnarray} 
$\chi(p,k)$ corresponds to the unamputated
vertex. Eq.(\ref{quark-target}) holds under the assumption of scaling
and finiteness in the Bjorken limit.

Among the hadrons, pseudoscalar mesons and, in particular the pion,
appear as the simplest and best understood states from a theoretical
viewpoint, since most of its properties can be explained in terms of
spontaneous chiral symmetry breaking. Its bound state properties are
not expected to depend crucially on confinement, at least for small
excitation energies. Furthermore, the pion structure functions have
been deduced from phenomenological QCD studies \cite{GRV98,SMRS92} and
there also exist lattice calculations of the first few moments
\cite{lat}. Thus, we do not expect to achieve a better theoretical
understanding of a DIS process than in the case of the pion. Hence,
structure functions of the pion have been computed several times in
different quark-loop models
\cite{SS93,FM94,DR95,SW95,JM97,BH99,WRG99,DT00,HRS01}.  The issue of
implementing gauge invariance and hence proper normalization of
$F(x,Q_0^2)$ is problematic and therefore requires some special care.

\section{Quark propagator and Ward-Takahashi identities}

I present here another model \cite{Ru01} for pion structure based in a
spectral representation of the quark propagator and chiral
Ward-Takahashi identities (WTI) for the flavoured vertex functions,
deduced from conservation of the vector current (CVC) and partial
conservation of the axial current (PCAC)
\begin{eqnarray}
 J^{\mu,a}_V (x) &=& \bar q(x) \gamma^\mu \frac{\tau_a}{2}
 q(x)\phantom{\gamma_5}  
 \qquad  
\partial_\mu J^{ \mu ,a}_V (x) = 0 
\label{cvc} 
\\ 
J^{\mu,a}_A (x)  &=& \bar q(x) \gamma^\mu \gamma_5 \frac{\tau_a}{2} q(x)  
 \qquad 
\partial_\mu J^{ \mu ,a}_A (x) = \bar M_0 \bar q (x) i \tau_a \gamma_5 q(x) 
\label{pcac} 
\end{eqnarray} 
with $\bar M_0$ the average up and down current quark masses. 

We assume a spectral representation for the quark propagator
\begin{eqnarray}
S(p) = \int d w \frac{ \rho( w )} {\slashchar{p} - w } = -i \int d^4 x
e^{i p \cdot x} \langle 0 \left| T \left\{ q(x) \bar q(0) \right\} \right| 0
\rangle
\label{spectral} 
\end{eqnarray}
The requirement of a physical Hilbert space would imply the positivity
of the spectral function $\rho(w)$, but we are not assuming this
here. Predictions based on Eq.~(\ref{spectral}) require a specific
form of $\rho(w)$ as input, but it turns out \cite{Ru01} that in the
chiral limit, $\bar M_0 \to 0 $ in Eq.~(\ref{pcac}), many properties
of interest are independent of $\rho(w)$. Thus we will focus on those
observables and consider the chiral limit. The vector and axial
unamputated vertex functions satisfy the WTI
\begin{eqnarray}
 (p'-p)_\mu  S(p') \Gamma_V^{\mu,a} (p', p) S(p) &=& \frac{\tau_a}{2}
 S(p) - S(p') \frac{\tau_a}{2} \\ (p'-p)_\mu S(p') \Gamma_A^{\mu,a} 
 (p', p) S(p) &=& S(p') \frac{\tau_a}{2} \gamma_5 + \gamma_5
 \frac{\tau_a}{2} S(p)
\end{eqnarray} 
The gauge technique, in the form introduced in QED \cite{DW77}
consists of writing {\it a solution} for the vector and axial vertices
in the form
\begin{eqnarray}
S(p') \Gamma_V^{\mu,a} (p', p) S(p) &=& \int d w \rho(w) \frac1{
\slashchar{p'} - w } \gamma^\mu \frac{\tau_a}{2} \frac1{
\slashchar{p} - w } \label{vector}\\ S(p') \Gamma_A^{\mu,a} (p', p) S(p)
 &=& \int d
w \rho(w) \frac1{\slashchar{p'} - w } \left( \gamma^\mu - \frac{2 w
q^\mu}{ q^2} \right) \gamma_5 \frac{\tau_a}{2} \frac1{
\slashchar{p} - w } \label{axial}
\end{eqnarray} 
with $q=p'-p$. An interesting consequence of the axial Ward identity
(\ref{axial}) is the presence of a massless pseudoscalar pole, which
is identified with the pion.

\section{Properties of the spectral function} 

Some properties on the spectral function $\rho(w)$ can be derived by
studying specific processes and comparing with known QCD results. For
instance, vacuum polarization can be computed closing the
vector-quark-quark vertex in Eq.~(\ref{vector}) 
\begin{eqnarray}
i \int d^4 x e^{-i p \cdot x} \langle 0
| T \left\{ J^{\rm em}_\mu (x) J^{\rm em}_\nu (0) \right\} | 0 \rangle =
\left( -g_{\mu\nu} + \frac{p_\mu p_\nu}{ p^2 } \right) \Pi^{\rm em} (p^2)  
\end{eqnarray} 
The $e^+ e^- \to {\rm hadrons} $ cross-section is proportional to the
imaginary part of the vacuum charge polarization
operator. Asymptotically, one has  
\begin{eqnarray}
\sigma( e^+ e^- \to {\rm hadrons}) & \to & 
\frac{4 \pi  \alpha^2}{ 3 s } \left(
\sum_i e^2_i \right) \int dw \rho(w)
\end{eqnarray} 
Thus, the proper QCD asymptotic result is obtained if 
\begin{eqnarray}
\int dw \rho(w) = 1  \label{norm}
\end{eqnarray} 
This condition is equivalent to impose that  $ \lim_{p\to \infty}
\slashchar{p} S(p) = 1 $.

%\subsection{Weak pion decay} 

The weak pion decay constant can be computed from the axial-axial
correlation function, since 
\begin{eqnarray}
\langle 0 \left| J_A^{\mu a} (x) \right| \pi_b (p) \rangle = i f_\pi
p_\mu \delta_{a,b} e^{i p \cdot x}
\end{eqnarray}
and then, inserting a complete set of eigenstates between the currents,
\begin{eqnarray}
 \int d^4 x e^{-i p \cdot x} \langle 0
| T \left\{ J_A^{\mu a} (x) J_A^{\nu b} (0) \right\} | 0 \rangle = i
f_\pi^2 \delta_{ab} \frac{p^\mu p^\nu}{p^2} + \dots
\end{eqnarray} 
where the dots indicate regular pieces in the limit $p^2 \to 0
$. Again, closing the axial-quark-quark vertex in Eq.~(\ref{axial}) to
make a quark loop, and going to the pion pole, $p^2 \to 0$, one gets
\begin{eqnarray}
f_\pi^2 = \frac{4 N_c}{(4\pi)^2 } \int dw \rho(w) \frac{1}{i} \int \frac{d^4 p}
{(2\pi)^4 } \frac{w^2}{( p^2 - w^2 )^2 }
\end{eqnarray} 
The momentum integral is logarithmically divergent. Thus, if  
\begin{eqnarray}   
\int dw w^2 \rho(w)=0 \qquad {\rm then} \qquad f_\pi^2 = 
\frac{4 N_c}{(4\pi)^2 } \int dw
w^2 (- \log w^2 ) \rho(w)
\label{quad} 
\end{eqnarray}
As we see, $\rho(w)$ cannot be a positive definite function,
regardless of its support. $f_\pi^2 $ is positive if $\rho(w)$ is
positive around $w=0$. Notice also that the first condition of
Eq.~(\ref{quad}) implies that the value of $f_\pi$ does not depend on
any particular scale used to make the logarithm dimensionless. If 
$\rho(w)$ had only one zero, the scale of $f_\pi$ would be set by it.   

A direct application of the previous conditions (\ref{norm}) and
(\ref{quad}) can be found in the calculation of the pion
electromagnetic form factor $F_\pi^{\rm em} (q^2 ) $ which complete
evaluation requires an explicitly knowledge of the spectral function
$\rho(w)$. Besides a proper normalization, $F_\pi^{\rm em}(0 )=1 $,
the mean squared radius becomes
\begin{eqnarray}
\langle r^2_\pi \rangle = -\frac16 \frac{d F}{ dq^2 } \Big|_{q^2 =0} = 
\frac{N_c}{4 \pi^2 f_\pi^2}  \int dw \rho(w) 
\end{eqnarray} 
which is the value obtained using the unregularized quark loop if the
normalization condition (\ref{norm}) is used. Thus, the pion has a
finite size in this framework.  

The amplitude for the (anomalous) neutral pion decay $ \pi^0 ( p ) \to
\gamma ( p_1 , \mu ) + \gamma (p_2 , \nu) $ can also be computed from
the triangle diagram yielding for on-shell massless pions
$p_1^2=p_2^2=0 $, the result
\begin{eqnarray}
M_{\mu \nu} (p_1,p_2) = \frac{N_c}{ 12 \pi^2 f_\pi } \epsilon_{\mu\nu
\alpha \beta}p_2^\alpha p_1^\beta \int dw \rho(w)
\end{eqnarray} 
which, again, coincides with the standard one if Eq.(\ref{norm}) is used. 

\section{Pion Structure Function} 

An advantage of our method is that calculations can be directly done
in Minkowski space and hence we may directly profit from LC
coordinates. For the evaluation of the structure function of the pion
the relevant vertex is constructed from two axial currents and is
defined
\begin{eqnarray}
(2\pi)^4 \delta (p'+q'-p-q)   && \chi_{AA}^{\mu,a;\nu,b} (p',q'; p,q )
\\ = i \int d^4 x d^4 x' d^4 y' d^4 y && \langle 0 | T
 \left\{ J_A^{\mu , a} (x) J_A^{\nu , b} (x') q(y) \bar q(y') \right\} 
| 0 \rangle  e^{i [ q' \cdot x' + p' \cdot y' - q \cdot x -p \cdot y ]}
\nonumber 
\end{eqnarray} 
and fulfills the WTI 
\begin{eqnarray}
i q^\mu  \chi_{AA}^{\mu,a;\nu,b} (p',q' ; p , q ) =  \epsilon_{bac} 
S(p') \Gamma_V^{\nu,c} (p',&p&) S(p) \\
+  \frac{\tau_a}{2} \gamma_5 S(p'-q) \Gamma_A^{\nu b} (p'-q,p) S(p) 
&+&  S(p') \Gamma_A^{\nu b} (p',p+q) S(p+q) \frac{\tau_a}{2} \gamma_5
\nonumber 
\end{eqnarray} 
Up to transverse pieces one gets the solution 
\begin{eqnarray}
&& \chi_{AA}^{\mu,a;\nu,b} (p',q'; p,q )  =   \int dw \rho(w) \frac1{
\slashchar{p}' -w} \Big\{ \frac{w {q'}^\nu q^\mu}{ {q'}^2 q^2} \delta_{ab}
\\ + &&   
\left( \gamma_\nu + \frac{2 w {q'}^2}{ {q'}^2} \right)  
\gamma_5 \frac{\tau_b}{ 2} \frac1{
\slashchar{p} + \slashchar{q} -w}  
\left( \gamma_\mu - \frac{2 w q^\mu}{ q^2} \right)
\gamma_5 \frac{\tau_a }{ 2} + {\rm crossed}   \Big\} \frac1{
\slashchar{p}' -w} 
\nonumber 
\end{eqnarray} 
At the pion poles, $q^2 , {q'}^2 \to 0 $, we get the $\pi q \to
\pi q $ unamputated amplitude in the forward direction ($q=q'$) 
\begin{eqnarray}
\chi_{AA}^{\mu,a;\nu,b} (p,q; p,q ) \to \frac{ q^\nu q^\mu}{ q^4
} f_\pi^2 \chi^{ba} (p,q)  
\end{eqnarray} 
and direct use of Eq.(\ref{quark-target}) together with the conditions
(\ref{norm}) and (\ref{quad}) yields the initial condition for the
quark distributions in $\pi^+$,
\begin{eqnarray}
u_\pi (x, Q_0^2 ) = \bar d_\pi (1-x, Q_0^2 ) = \theta(x) \theta(1-x) 
\label{result} 
\end{eqnarray}
This result has been derived in the chiral limit, is independent of 
the spectral function $ \rho(w) $ and has proper support and
normalization. By construction, Eq.(\ref{result}) is consistent
with chiral symmetry, although is not necessarily a consequence of
it. The result has previously been obtained by several means within
the NJL model either using Pauli-Villars regularization
\cite{DR95,WRG99} on the virtual Compton amplitude
(\ref{virtual-compton}) or imposing a transverse cut-off \cite{BH99}
upon the quark-target amplitude (\ref{quark-target}). Within the
present approach it can be shown that the forward Compton scattering
amplitude (\ref{virtual-compton}) in the Bjorken limit yields the
finite result presented here \cite{Ru01} based on
Eq.(\ref{quark-target}).  To really appreciate this point, let us
mention that Eq.(\ref{result}) disagrees with other NJL calculations,
due to the use of different regularizations. If
Eq.(\ref{virtual-compton}) is used with a four-dimensional cut-off
\cite{SS93} or  Eq.(\ref{quark-target}) is used with
Lepage-Brodsky regularization \cite{BH99}, different shapes for the
quark distributions are obtained. The null-plane \cite{FM94}, NJL
model \cite{SS93} and spectator model \cite{JM97} calculations also
produce different results. In all cases, the use of momentum dependent
form factors or non-gauge invariant regularizations make the
connection between Eq.(\ref{virtual-compton}) and
Eq.(\ref{quark-target}) doubtful and, furthermore, spoil
normalization. The results based on a quark loop with momentum
dependent quark masses \cite{DT00,HRS01} seem to produce a non-constant distribution.

\begin{figure}[t]
\begin{center}
\epsfig{figure=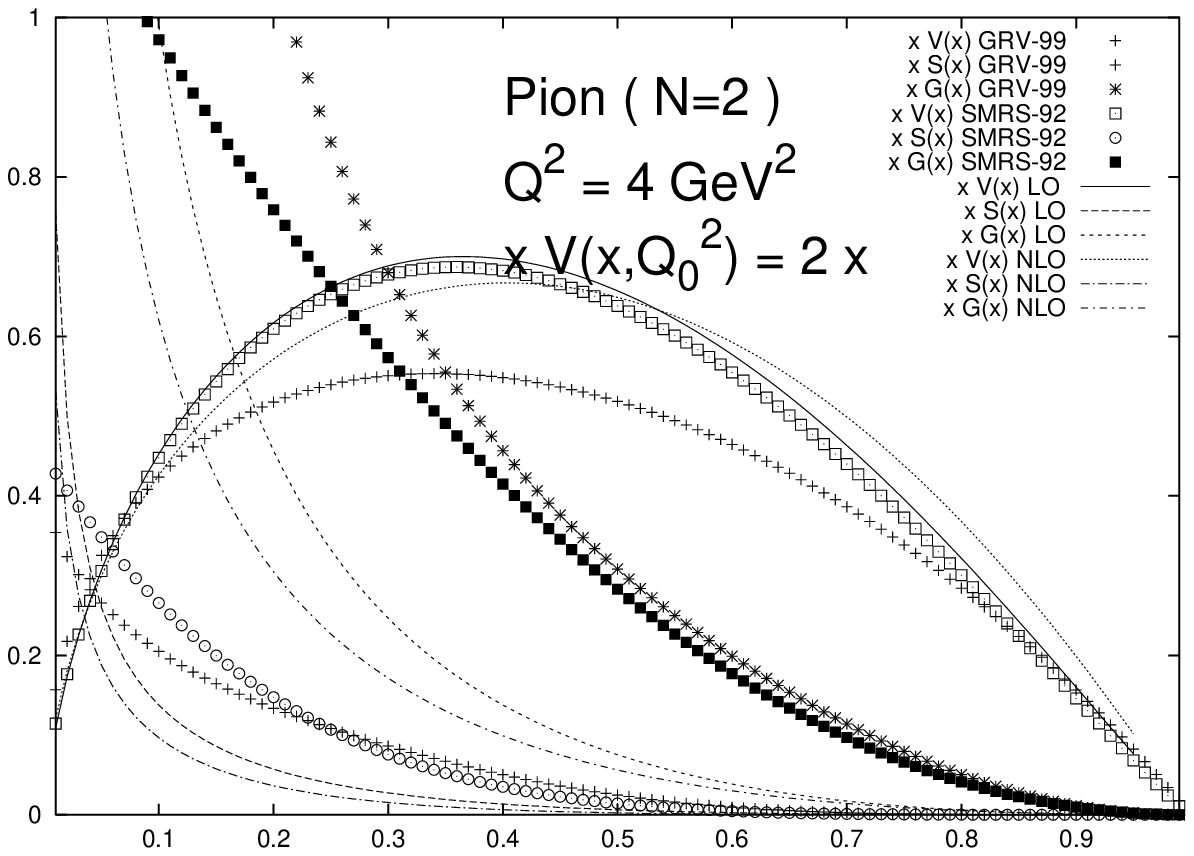,height=6.8cm,width=5.8cm}\epsfig{figure=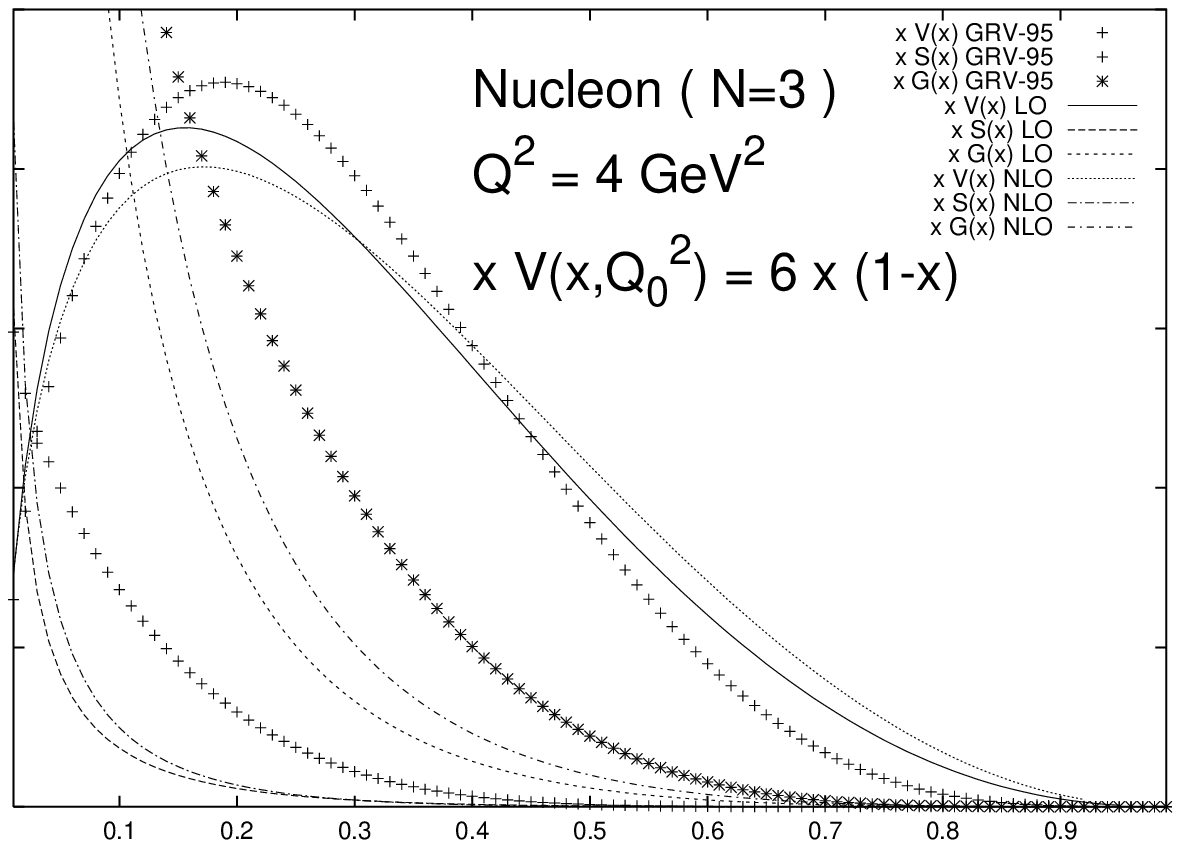,height=6.8cm,width=5.8cm}
\end{center}
\caption{Valence, gluon and sea distributions in the pion (left) and
the nucleon (right) at \protect{$ Q^4 = 4 {\rm GeV}^2 $} compared with
phenomenological analysis for the pion SMRS92 ${}^7$, GRV99 ${}^8$ and
for the nucleon GRV95 ${}^3$. We take \protect{$ \langle x V
\rangle_\pi = 0.47 $} and \protect{$\langle x V \rangle_N = 0.40$ at $
Q^2 = 4 {\rm GeV}^2 $}.  The sea and gluon distributions fulfill
\protect{$ S(x,Q_0^2)=G(x,Q_0^2)=0 $}. }
\label{fig:dist}
\end{figure}
\section{QCD evolution} 
Eq.(\ref{result}) can also be understood \cite{Ru98b} in terms of
phase space arguments and point couplings (i.e., constant matrix
elements) which for $N$ constituents gives 
\begin{eqnarray} 
x V(x,Q_0^2 ) = N (N-1) x (1-x)^{N-2} \qquad \langle x V\rangle \equiv
\int_0^1 dx V(x,Q_0^2) x = 1
\label{eq:initial} 
\end{eqnarray} 
Of course, it is tempting to look at the nucleon case $N=3$, although
we obviously expect mass corrections to be more important both in the
momentum dependence of the $Nqqq$ matrix element and in the phase
space, and, in addition we do not have as much theoretical support as
in the pion case. As usual, we evolve Eq.~(\ref{eq:initial}) at LO and
NLO exactly implementing the group properties mentioned after
Eq.(\ref{evol}) and employed to look for low energy unitarity limits
\cite{Ru98a}. Assuming vanishing initial sea and gluon distributions,
the result for the dynamically generated valence,sea and gluon
distributions for a hadron with $N=2$ and $N=3$ constituents compared
with known parameterizations for the pion \cite{SMRS92,GRV99} and the
nucleon \cite{GRV95} respectively can be seen in
Fig~(\ref{fig:dist}). The agreement with the phenomenological valence
distributions is good. Actually, since normalization of the valence
distributions is preserved by evolution and $Q_0^2$ is determined by
constraining the valence momentum fraction at $Q^2=4 {\rm GeV}^2 $ to
the phenomenological ones, $ \langle x V \rangle_\pi=0.47$ and $
\langle x V \rangle_N = 0.40$ the bulk of the distribution is
reproduced.  Sea and gluon distributions do not obey such strong
constraints and provide more information regarding the quality of a
model. As we see, they are not as good. Nevertheless, a more definite
statement might be made if uncertainties in the distributions were
considered on the QCD evolution side \cite{Derrick}. The estimate of
systematic errors in the model, i.e. corrections to the initial
condition, remains yet a difficult and open problem.

%\newpage
\section*{Acknowledgments}
I warmly thank the organizers for the invitation.  I thank
R. Delbourgo for discussions at CSSM where this work was
started. Support from DGES (Spain) Project PB98-1367 and by the Junta
de Andaluc\'\i a is acknowledged.

\end{document}